\documentclass{llncs}
\input{preambule.sty}

\title{Joint Total Variation ESTATICS\\for Robust Multi-Parameter Mapping}

\author{%
Ya\"el Balbastre\inst{1} \and
Mikael Brudfors\inst{1} \and
Michela Azzarito\inst{2} \and
Christian Lambert\inst{1} \and
Martina F. Callaghan\inst{1} \and
John Ashburner\inst{1}}

\institute{
Wellcome Centre for Human Neuroimaging,\\
Queen Square Institute of Neurology, %
University College London, %
London, UK.%
\and%
Spinal Cord Injury Center Balgrist,\\
University Hospital Zurich, %
University of Zurich, %
Switzerland.}

\begin{document}

\maketitle

\begin{abstract}
Quantitative magnetic resonance imaging (qMRI) derives tissue-specific parameters -- such as the apparent transverse relaxation rate $R_2^\star$, the longitudinal relaxation rate $R_1$ and the magnetisation transfer saturation -- that can be compared across sites and scanners and carry important information about the underlying microstructure. The multi-parameter mapping (MPM) protocol takes advantage of multi-echo acquisitions with variable flip angles to extract these parameters in a clinically acceptable scan time. In this context, ESTATICS performs a joint loglinear fit of multiple echo series to extract $R_2^\star$ and multiple extrapolated intercepts, thereby improving robustness to motion and decreasing the variance of the estimators. In this paper, we extend this model in two ways: (1) by introducing a joint total variation (JTV) prior on the intercepts and decay, and (2) by deriving a nonlinear maximum \emph{a posteriori} estimate. We evaluated the proposed algorithm  by predicting left-out echoes in a rich single-subject dataset. In this validation, we outperformed other state-of-the-art methods and additionally showed that the proposed approach greatly reduces the variance of the estimated maps, without introducing bias.
\end{abstract}


\section{Introduction}

The magnetic resonance imaging (MRI) signal is governed by a number of tissue-specific parameters. While many common MR sequences only aim to maximise the contrast between tissues of interest, the field of quantitative MRI (qMRI) is concerned with the extraction of the original parameters \cite{Tofts2003}. This interest stems from the fundamental relationship that exists between the magnetic parameters and the tissue microstructure: the longitudinal relaxation rate $R_1=1/T_1$ is sensitive to myelin content \cite{Sigalovsky2006,Dick2012,Sereno2013}; the apparent transverse relaxation rate $R_2^\star=1/T_2^\star$ can be used to probe iron content \cite{Ordidge1994,Ogg1999,Hasan2012}; the magnetization-transfer saturation (MT\textsubscript{sat}) indicates the proportion of protons bound to macromolecules (in contrast to free water) and offers another metric to investigate myelin loss \cite{Tofts2003b,Helms2008b}. Furthermore, qMRI allows many of the scanner- and centre-specific effects to be factored out, making measures more comparable across sites \cite{Tofts2006,Deoni2008,Bauer2010,Weiskopf2013a}. In this context, the multi-parameter mapping (MPM) protocol was developed at 3 Tesla to allow the quantification of $R_1$, $R_2^\star$, MT\textsubscript{sat} and the proton density (PD) at high resolutions (0.8 or 1 mm) and in a clinically acceptable scan time of 25 mins \cite{Helms2008b,Weiskopf2013a}. However, to reach these values, compromises must be made so that the signal-to-noise ratio (SNR) suffers, making the parameter maps noisy; Papp et al. \cite{Papp2016a} found a scan-rescan root mean squared error of about 7.5\% for $R_1$ at 1mm, in the absence of inter-scan movement. Smoothing can be used to improve SNR, but at the cost of lower spatial specificity.

Denoising methods aim to separate signal from noise. They take advantage of the fact that signal and noise have intrinsically different spatial profiles: the noise is spatially independent and often has a characteristic distribution while the signal is highly structured. Denoising methods originate from partial differential equations, adaptive filtering, variational optimisation or Markov random fields, and many connections exist between them. Two main families emerge:
\begin{enumerate}
    \item Optimisation of an energy:
    $\textstyle\hat{Y} = \argmin_Y \mathcal{E}_1\left(X - \mathcal{A}(Y)\right) + \mathcal{E}_2\left(\mathcal{G}(Y)\right),$\\
    where $X$ is the observed data, $Y$ is the unknown noise-free data, $\mathcal{A}$ is an arbitrary \emph{forward} transformation (\emph{e.g.}, spatial transformation, downsampling, smoothing) mapping from the reconstructed to the observed data and $\mathcal{G}$ is a linear transformation (\emph{e.g.}, spatial gradients, Fourier transform, wavelet transform) that extracts features of interest from the reconstruction.
    \item Application of an adaptive nonlocal filter: 
    $\textstyle\hat{Y}_i = \sum_{j\in\mathcal{N}_i} w\left(\mathcal{P}_i(X),\mathcal{P}_j(X)\right) X_j,$\\
    where the reconstruction of a given voxel $i$ is a weighted average all observed voxels $j$ in a given (possibly infinite) neighbourhood $\mathcal{N}_i$, with weights reflecting similarity between patches centred about these voxels.
\end{enumerate}
For the first family of methods, it was found that the denoising effect is stronger when $\mathcal{E}_2$ is an absolute norm (or sum of), rather than a squared norm, because the solution is implicitly sparse in the feature domain \cite{Bach2011}. This family of methods include total variation (TV) regularisation \cite{Rudin1992a} and wavelet soft-thresholding \cite{Donoho1995}. The second family also leverages sparsity in the form of redundancy in the spatial domain; that is, the dictionary of patches necessary to reconstruct the noise-free images is smaller than the actual number of patches in the image. Several such methods have been developed specifically for MRI, with the aim of finding an optimal, voxel-wise weighting based on the noise distribution \cite{Coupe2008c,Manjon2010d,Coupe2012,Manjon2012b}.

Optimisation methods can naturally be interpreted as a maximum \emph{a posteriori} (MAP) solution in a generative model, which eases its interpretation and extension. This feature is especially important for MPMs, where we possess a well-defined (nonlinear) forward function and wish to regularise a small number of maps. In this paper, we use the ESTATICS forward model \cite{Weiskopf2014a}, which assumes a shared $R_2^\star$ decay across contrasts, with a joint total variation (JTV) prior.  JTV \cite{Sapiro1996} is an extension of TV to multi-channel images, where the absolute norm is defined across channels, introducing an implicit correlation between them. TV and JTV have been used before in MR reconstruction (\emph{e.g.}, in compressed-sensing \cite{Huang2012}, quantitative susceptibility mapping \cite{Liu2011}, super-resolution\cite{Brudfors2018}). JTV is perfectly suited for modelling the multiple contrasts in MPMs and increases the power of the implicit edge-detection problem. However, a challenge stems from the nonlinear forward model that makes the optimisation problem nonconvex.

Our implementation uses a quadratic upper bound of the JTV functional and the surrogate problem is solved using second-order optimisation. Positive-definiteness of the Hessian is enforced by the use of Fisher's scoring, and the quadratic problem is efficiently solved using a mixture of multi-grid relaxation and conjugate gradient. We used a unique dataset -- five repeats of the MPM protocol acquired, within a single session, on a healthy subject -- to validate the proposed method. Our method was compared to two variants of ESTATICS: loglinear \cite{Weiskopf2014a} and Tikhonov-regularised. We also compared it with the adaptive optimized nonlocal means (AONLM) method \cite{Manjon2010d}, which is recommended for accelerated MR images (as is the case in our validation data). In that case, individual echoes were denoised using AONLM, and maps were reconstructed with the loglinear variant of ESTATICS. In our validation, JTV performed consistently better than all other methods.

\section{Methods}

\noindent\textbf{Spoiled Gradient Echo.} The MPM protocol uses a multi-echo spoiled gradient-echo (SGE) sequence with variable flip angles to generate weighted images. The signal follows the equation:
\begin{align}
    \textstyle
    S(\alpha,T_R,T_E) = S_0(\alpha,T_R)\exp(-T_E R_2^\star)~,
    \label{eq:sge}
\end{align}
where $\alpha$ is the nominal flip angle, $T_R$ is the repetition time and $T_E$ is the echo time. PD and T1 weighting are obtained by using two different flip angles, while MT weighting is obtained by playing a specific off-resonance pulse beforehand. If all three intercepts $S_0$ are known, rational approximations can be used to compute $R_1$ and MT$_{\mathrm{sat}}$ maps \cite{Helms2008b,Helms2008c}.\\\vspace{-0.2cm}

\noindent\textbf{ESTATICS.} ESTATICS aims to recover the decay rate $R_2^\star$ and the different intercepts from \eqref{eq:sge}. We therefore write each weighted signal (indexed by $c$) as:
\begin{align}
    S(c,T_E) = \exp(\theta_c - T_ER_2^\star)~,~~ \mathrm{with}~~ \theta_c = \ln S_{0c}~.
\end{align}
At the SNR levels obtained in practice ($> 3$), the noise of the log-transformed data is approximately Gaussian (although with a variance that scales with signal amplitude). Therefore, in each voxel, a least-squares fit can be used to estimate $R_2^\star$ and the log-intercepts $S_c$ from the log-transformed acquired images.\\\vspace{-0.2cm}

\noindent\textbf{Regularised ESTATICS.} Regularisation cannot be easily introduced in logarithmic space because, there, the noise variance depends on the signal amplitude, which is unknown. Instead, we derive a full generative model. Let us assume that all weighted volumes are aligned and acquired on the same grid. Let us define the image acquired at a given echo time $t$ with contrast $c$ as $\vec{s}_{c,t} \in \mathbb{R}^{I}$ (where $I$ is the number of voxels). Let $\vec{\theta}_c \in \mathbb{R}^I$ be the log-intercept with contrast $c$ and let $\vec{r} \in \mathbb{R}^{I}$ be the $R_2^\star$ map. Assuming stationary Gaussian noise, we get the conditional probability:
\begin{align}
    \textstyle
    \p{\vec{s}_{c,t}}{\vec{\theta}_c, \vec{r}} = \N{\vec{s}_{c,t}}{\tilde{\vec{s}}_{c,t},~ \sigma_c^2\vec{I}} ~,~~
    \tilde{\vec{s}}_{c,t} = \exp(\vec{\theta}_c - t\vec{r}) ~.
\label{eq:dataterm}
\end{align}
The regularisation takes the form of a joint prior probability distribution over $\vec{\Theta} = \left[\vec{\theta}_1,~\cdots, ~\vec{\theta}_C,~\vec{r}\right]$. For JTV, we get:
\begin{align}
    \textstyle
    \p{\vec{\Theta}} \propto \prod_i \exp\left(-\sqrt{\sum_{c=1}^{C+1} \lambda_c \vec{\theta}_{c}\T\vec{G}_i\T\vec{G}_i\vec{\theta}_{c}}\right) ~,
\end{align}
where $\vec{G}_i$ extracts all forward and backward finite-differences at the $i$-th voxel and $\lambda_c$ is a contrast-specific regularisation factor. The MAP solution can be found by maximising the joint loglikelihood with respect to the parameter maps.\\\vspace{-0.2cm}

\noindent\textbf{Quadratic Bound.} The exponent in the prior term can be written as the minimum of a quadratic function \cite{Daubechies2010,Bach2011}:
\begin{align}
    {\textstyle
    \sqrt{\sum_c \lambda_c \vec{\theta}_{c}\T\vec{G}_i\T\vec{G}_i\vec{\theta}_{c}}}
    = \min_{w_i > 0} \left\{ \frac{w_i}{2} + \frac{1}{2w_i} \sum_c \lambda_c \vec{\theta}_{c}\T\vec{G}_i\T\vec{G}_i\vec{\theta}_{c} \right\} .
    \label{eq:bound}
\end{align}
When the weight map $\vec{w}$ is fixed, the bound can be seen as a Tikhonov prior with nonstationary regularisation, which is a quadratic prior that factorises across channels. Therefore, the between-channel correlations induces by the JTV prior are entirely captured by the weights. Conversely, when the parameter maps are fixed, the weights can be updated in closed-form:
\begin{align}
    \textstyle
    w_i = \sqrt{\sum_c \lambda_c \vec{\theta}_{c}\T\vec{G}_i\T\vec{G}_i\vec{\theta}_{c}} ~.
\end{align}
The quadratic term in \eqref{eq:bound} can be written as $\lambda_c\vec{\theta}_c\T\vec{L}\vec{\theta}_c$, with $\vec{L} = \sum_i \frac{1}{w_i} \vec{G}_i\T\vec{G}_i$. 

In the following sections, we will write the full (bounded) model negative loglikelihood as $\mathcal{L}$ and keep only terms that depend on $\vec{\Theta}$, so that:
\begin{align}
    \mathcal{L} = \sum_{c,t} \mathcal{L}^{\mathrm{d}}_{c,t} + \mathcal{L}^{\mathrm{p}},~
    \mathcal{L}^{\mathrm{d}}_{c,t} \cequal \frac{1}{2\sigma_c^2}\lVert\vec{s}_{c,t}-\tilde{\vec{s}}_{c,t}\rVert^2,~
    \mathcal{L}^{\mathrm{p}} \cequal \frac{1}{2} \sum_c \vec{\theta}_{c}\T\vec{L}_c\vec{\theta}_{c}.
\end{align}

\noindent\textbf{Fisher's Scoring.} The data term \eqref{eq:dataterm} does not always have a positive semi-definite Hessian (it is not convex). There is, however, a unique optimum. Here, to ensure that the conditioning matrix that is used in the Newton-Raphson iteration has the correct curvature, we take the expectation of the true Hessian, which is equivalent to setting the residuals to zero -- a method known as Fisher's scoring. The Hessian of $\mathcal{L}^{\mathrm{d}}_{c,t}$ with respect to the $c$-th intercept and $R_2^\star$ map then becomes:
\begin{align}
    \vec{H}^{\mathrm{d}}_{c,t} &{}= \frac{1}{\sigma^2} \mathrm{diag}\left(\tilde{\vec{s}}_{c,t}\right)
     \otimes 
    \left[\begin{array}{ccc}1 & & \text{-}t \\ \text{-}t & & t^2\end{array}\right]
    ~.
\end{align}

\noindent\textbf{Misaligned Volumes.}
Motion can occur between the acquisitions of the different weighted volumes. Here, volumes are systematically co-registered using a skull-stripped and bias-corrected version of the first echo of each volume. However, rather than reslicing the volumes onto the same space, which modifies the original intensities, misalignment is handled within the model. To this end, equation \eqref{eq:dataterm} is modified to include the projection of each parameter map onto native space, such that
$\tilde{\vec{s}}_{c,t} = \exp(\vec{\Psi}_c\vec{\theta}_c - t\vec{\Psi}_c\vec{r})$,
where $\vec{\Psi}_c$ encodes trilinear interpolation and sampling with respect to the pre-estimated rigid transformation. The Hessian of the data term becomes $\vec{\Psi}_c\T\vec{H}^{\mathrm{d}}_{c,t}\vec{\Psi}_c$, which is nonsparse. However, an approximate Hessian can be derived \cite{Ashburner2018}, so that:
\begin{align}
    \vec{H}^{\mathrm{d}}_{c,t} \approx \frac{1}{\sigma^2} \mathrm{diag}\left(\vec{\Psi}_c\T\tilde{\vec{s}}_{c,t}\right)
     \otimes 
    \left[\begin{array}{ccc}1 & & \text{-}t \\ \text{-}t & & t^2\end{array}\right]
    ~.
\end{align}
Since all elements of $\tilde{\vec{s}}_{c,t}$ are strictly positive, this Hessian is ensured to be more positive-definite than the true Hessian in the Löwner ordering sense.\\\vspace{-0.2cm}

\noindent\textbf{Newton-Raphson.} The Hessian of the joint negative log-likelihood becomes:
\begin{align}
    \vec{H} = \vec{H}^{\mathrm{d}} + \vec{L} \otimes \mathrm{diag}\left(\vec{\lambda}\right) ~.
\end{align}
Each Newton-Raphson iteration involves solving for $\vec{H}^{-1}\vec{g}$, where $\vec{g}$ is the gradient. Since the Hessian is positive-definite, the method of conjugate gradients (CG) can be used to solve the linear system. CG, however, converges quite slowly. Instead, we first approximate the regularisation Hessian $\vec{L}$ as \newline$\tilde{\vec{L}}=\frac{1}{\min\left(\vec{w}\right)}\sum_i \vec{G}_i\T\vec{G}_i$, which is more positive-definite than $\vec{L}$. Solving this substitute system therefore ensures that the objective function improves. Since $\vec{H}^{\mathrm{d}}$ is an easily invertible block-diagonal matrix, the system can be solved efficiently using a multi-grid approach \cite{Press2007}. This result is then used as a warm start for CG. Note that preconditioners have been shown to improve CG convergence rates \cite{Chen2018a,Xu2015a}, at the cost of slowing down each iteration. Here, we have made the choice of performing numerous cheap CG iterations rather than using an expensive preconditioner.

\section{Validation}

\textbf{Dataset.} A single participant was scanned five times in a single session with the 0.8 mm MPM protocol, whose parameters are provided in table \ref{table:mpm}. Furthermore, in order to correct for flip angles nonhomogeneity, a map of the $B_1^+$ field was reconstructed from stimulated and spin echo 3D EPI images \cite{Lutti2010a}.\\\vspace{-0.2cm} 

\begin{table}
\caption{Sequence parameters of the MPM protocol. The MTw sequence has an off-resonance prepulse (PP): 220$^\circ$, 4ms duration, 2kHz off-resonance.}
\begin{widetable}{\textwidth}{r | cccccc}
    \toprule
     & \textbf{FA} & \textbf{TR} & \textbf{TE} & \textbf{Matrix} & \textbf{FOV} & \textbf{PP}\\
    \midrule
    \textbf{T1w} & 21$^{\circ}$ & 25ms & $[1..8] \times 2.3$ms & $320 \times 280 \times 224$ & $256 \times 224 \times 179.2$ mm$^3$ & \\
    \textbf{PDw} & 6$^{\circ}$ & 25ms & $[1..8] \times 2.3$ms & $320 \times 280 \times 224$ & $256 \times 224 \times 179.2$ mm$^3$ & \\
    \textbf{MTw} & 6$^{\circ}$ & 25ms & $[1..6] \times$ 2.3ms & $320 \times 280 \times 224$ & $256 \times 224 \times 179.2$ mm$^3$ & \checkmark\\
    \bottomrule
\end{widetable}
\label{table:mpm}
\end{table}

\noindent\textbf{Evaluated Methods.} Three ESTATICS methods were evaluated: a simple loglinear fit (LOG) \cite{Weiskopf2014a}, a nonlinear fit with Tikhonov regularisation (TKH) and a nonlinear fit with joint total variation regularisation (JTV). Additionally, all echoes were denoised using the adaptive nonlocal means method (AONLM) \cite{Manjon2010d} before performing a loglinear fit. The loglinear and nonlinear ESTATICS fit were all implemented in the same framework, allowing for misalignment between volumes. Regularised ESTATICS uses estimates of the noise variance within each volume, obtained by fitting a two-class Rice mixture to the first echo of each series. Regularised ESTATICS possesses two regularisation factors, one for each intercept and one for the $R_2^\star$ decay, while AONLM has one regularisation factor. These hyper-parameters were optimised by cross-validation (CV) on the first repeat of the MPM protocol.\\\vspace{-0.2cm}

\noindent\textbf{Leave-One-Echo-Out.} Validating denoising methods is challenging in the absence of a ground truth. Classically, one would compute similarity metrics, such as the root mean squared error, the peak signal-to-noise ratio, or the structural similarity index between the denoised images and noise-free references. However, in MR, such references are not artefact free: they are still relatively noisy and, as they require longer sequences, more prone to motion artefacts. A better solution is to use cross-validation, as the forward model can be exploited to predict echoes that were left out when inferring the unknown parameters. We fitted each method to each MPM repeat, while leaving one of the acquired echoes out. The fitted model was then used to predict the missing echo. The quality of these predictions was scored by computing the Rice loglikelihood of the true echo conditioned on the predicted echo within the grey matter (GM), white matter (WM) and cerebro-spinal fluid (CSF). An aggregate score was also computed in the parenchyma (GM+WM). As different echoes or contrasts are not similarly difficult to predict, Z-scores  were computed by normalising across repeats, contrasts and left-out echoes. This CV was applied to the first repeat to determine optimal regularisation parameters. We found $\beta=0.4$ without Rice-specific noise estimation to work better for AONLM, while for JTV we found $\lambda_1=5 \times 10^3$ for the intercepts and $\lambda_2=10$ for the decay (in $s^{-1}$) to be optimal.\\\vspace{-0.2cm}

\noindent\textbf{Quantitative Maps.} Rational approximations of the signal equations \cite{Helms2008b,Helms2008c} were used to compute $R_1$ and MT$_{\mathrm{sat}}$ maps from the fitted intercepts. The distribution of these quantitative parameters was computed within the GM and WM. Furthermore, standard deviation (S.D.) maps across runs were computed for each method.

\section{Results}

\begin{figure}
    \centering
    \includegraphics[width=\textwidth]{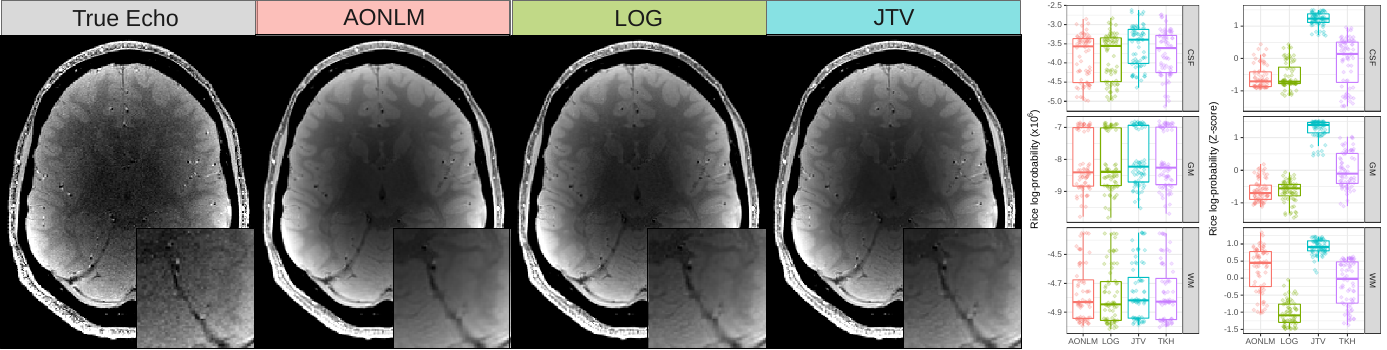}
    \caption{Leave-one-echo out prediction. Left: the true PDw echo at $T_E=9.7$ms from the 5th repeat and three predicted images. Right: boxplots of the Rice log-pdf and corresponding Z-score computed for each method within GM, WM and CSF masks. }
    \label{fig:predict}
\end{figure}

\textbf{Leave-One-Echo-Out.} The distribution of Rice loglikelihoods and Z-scores for each methods are depicted in Fig. \ref{fig:predict} in the form of Tukey's boxplots. In the parenchyma, JTV obtained the best score (mean log-likelihood: -$9.15\times 10^6$, mean Z-score: $1.19$) followed by TKH (-$9.26\times 10^6$ and -$0.05$), AONLM (-$9.34\times 10^6$ and -$0.41$) and LOG (-$9.35\times 10^6$ and -$0.72$).  As some echoes are harder to predict than others (typically, early echoes because their absence impacts the estimator of the intercept the most) the log-pdf has quite a high variance. However, Z-scores show that, for each echo, JTV does consistently better than all other methods. As can be seen in Fig. \ref{fig:predict}, JTV is particularly good at preserving vessels.

\begin{figure}
    \centering
    \includegraphics[width=0.8\textwidth]{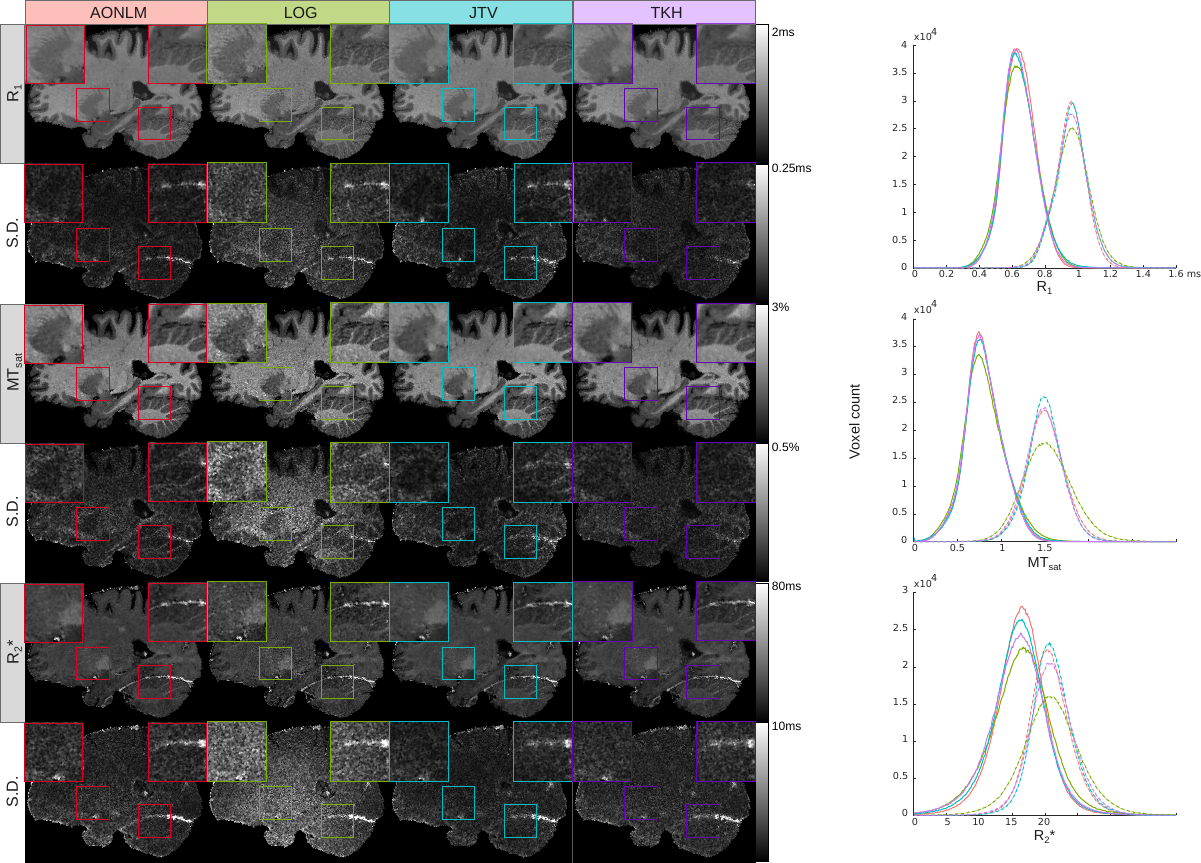}
    \caption{Quantitative maps. Left: example $R_1$, MT$_{\mathrm{sat}}$ and $R_2^\star$ maps obtained with each method, and standard deviation (S.D.) maps computed across runs. Right: mean intensity histograms computed within the GM (plain) and WM (dotted) masks.}
    \label{fig:map}
\end{figure}

\noindent\textbf{Quantitative Maps.} $R_1$, MT$_{\mathrm{sat}}$ and $R_2^\star$ maps reconstructed with each method are shown in Fig. \ref{fig:map}, along with mean intensity histograms within GM and WM. Note that these maps are displayed for qualitative purposes; low standard deviations are biased toward over-regularised methods and do not necessarily indicate a better predictive performance. It is evident from the histograms that all denoising methods sharpen the peaks without introducing apparent bias. It can be seen that JTV has lower variance than AONLM in the centre of the brain and higher in the periphery. This is because in our probabilistic setting, there is a natural balance between the prior and the quality of the data. In the centre of the brain, the SNR is lower than in the periphery, which gives more weight to the prior and induces a smoother estimate. The mean standard deviation of AONLM, LOG, JTV and TKH is respectively 9.5, 11.5, 11.5, 9.9 $\times 10^{-3}$ in the GM and 8.6, 12, 9.6, 10 $\times 10^{-3}$ in the WM for $R_1$, 15, 2, 17, 20 in the GM and 11, 20, 10, 13 in the WM for $R_2^\star$, and 4.6, 5.8, 5.1, 4.5 $\times 10^{-2}$ in the GM and 4.9, 8.2, 4.3, 4.7 $\times 10^{-2}$ in the WM for MT$_{\mathrm{sat}}$. Once again, variance is reduced by all denoising methods compared to the nonregularised loglinear fit. Again, a lower variance does not necessarily indicate a better (predictive) fit, which can only be assessed by the CV approach proposed above.

\section{Discussion \& Conclusion}

In this paper, we introduce a robust, regularisation-based reconstruction method for quantitative MR mapping. The joint total variation prior takes advantage of the multiple MPM contrasts to increase its edge-detection power. Our approach was validated using an unbiased CV scheme, where it compared favourably over other methods, including a state-of-the-art MR denoising technique. It was shown to reduce the variance of the estimated parametric maps over non-regularised approaches, which should translate into increased power in subsequent cross-sectional or longitudinal voxel-wise studies. The use of a well-defined forward model opens the door to multiple extensions: the projection operator could be modified to include other components of the imaging process such as non-homogeneous receive fields or gridding, which would allow for joint reconstruction and super-resolution; parameters that are currently fixed \emph{a priori}, such as the rigid matrices, could be given prior distribution and be optimised in an interleaved fashion; non-linear deformations could be included to account for changes in the neck position between scans; finally, the forward model could be unfolded further so that parameter maps are directly fitted, rather than weighted intercepts. An integrated approach like this one could furthermore include and optimise for other components of the imaging process, such as non-homogeneous transmit fields. In terms of optimisation, our approach should benefit from advances in conjugate gradient preconditioning or other solvers for large linear systems. Alternatively, JTV could be replaced with a patch-based prior. Nonlocal filters are extremely efficient at denoising tasks and could be cast in a generative probabilistic framework, where images are built using a dictionary of patches \cite{Lebrun2013}. Variational Bayes can then be used to alternatively estimate the dictionary (shared across a neighbourhood, a whole image, or even across subjects) and the reconstruction weights.

\subsubsection{Acknowledgements:} 
YB, MFC and JA were funded by the MRC and Spinal Research Charity through the ERA-NET Neuron joint call (MR/R000050/1). 
MB and JA were funded by the EU Human Brain Project’s Grant Agreement No 785907 (SGA2).
MB was funded by the EPSRC-funded UCL Centre for Doctoral Training in Medical Imaging (EP/L016478/1) and the Department of Health NIHR-funded Biomedical Research Centre at University College London Hospitals. 
CL is supported by an MRC Clinician Scientist award (MR/R006504/1).
The Wellcome Centre for Human Neuroimaging is supported by core funding from the Wellcome [203147/Z/16/Z].

\bibliographystyle{splncs04}
\bibliography{main}

\end{document}